\pdfoutput=1
\documentclass[usenatbib]{mn2e}
\usepackage{apjfonts}
\usepackage{amssymb}
\usepackage{amsmath}
\usepackage{ctable}
\usepackage{url}
\usepackage{fixltx2e} 

\newcommand{\be}{\begin{equation}}
\newcommand{\ee}{\end{equation}}

\newcommand{\msun}{M_{\sun}}

\newcommand{\paperone}{Paper {\small I}}
\newcommand{\papertwo}{Paper {\small II}}
\newcommand{\paperthree}{Paper {\small III}}

\newcommand{\movieurl}{\url{https://www.cfa.harvard.edu/~phopkins/Site/Research.html}}

\newcommand\plotonesize[2]
 {\centering \leavevmode \includegraphics[width={#2\columnwidth}]{#1}}
\newcommand{\plotsidesize}[2]
 {\centering \leavevmode \includegraphics[width={#2\textwidth}]{#1}}
\newcommand{\acknowledgments}{\begin{small}\section*{Acknowledgments}\end{small}}
\newcommand\altaffilmark[1]{$^{#1}$}
\newcommand\altaffiltext[1]{$^{#1}$}
\title[Feedback \&\ Clump Coalescence]{Stellar Feedback \&\ Bulge Formation in Clumpy Disks
\vspace{-0.5cm}}

\author[Hopkins et al.]{
\parbox[t]{\textwidth}{ 
Philip F. Hopkins\altaffilmark{1}\thanks{Einstein Fellow. E-mail:phopkins@astro.berkeley.edu},
Dusan Kere\v{s}\altaffilmark{1,2}, 
Norman Murray\altaffilmark{3,4}, \\
Eliot Quataert\altaffilmark{1}, \&\ 
Lars Hernquist\altaffilmark{5}
} 
\vspace*{6pt} \\
\altaffiltext{1}{Department of Astronomy and Theoretical Astrophysics Center, University of California
  Berkeley, Berkeley, CA 94720} \\
  \altaffiltext{2}{Hubble Fellow} \\
  \altaffiltext{3}{Canadian Institute for Theoretical Astrophysics, 
60 St.\ George Street, University of Toronto, ON M5S 3H8, Canada} \\
\altaffiltext{4}{Canada Research Chair in Astrophysics} \\
\altaffiltext{5}{Harvard-Smithsonian Center for Astrophysics, 60 
Garden Street, Cambridge, MA 02138, USA}
\vspace{-0.5cm}
}

\date{Submitted to MNRAS, November, 2011\vspace{-0.6cm}}
\begin{document}
\maketitle
\label{firstpage}

\begin{abstract}

We use numerical simulations of isolated galaxies to study the effects of stellar feedback  on the formation and evolution of giant star-forming gas ``clumps'' in high-redshift, gas-rich galaxies. Such galactic disks are unstable to the formation of bound gas-rich clumps whose properties initially depend only on global disk properties, not the microphysics of feedback. In simulations without stellar feedback, clumps turn an order-unity fraction of their mass into stars and sink to the center, forming a large bulge and kicking most of the stars out into a much more extended stellar envelope.  By contrast, strong radiative stellar feedback disrupts even the most massive clumps after they turn $\sim10-20\%$ of their mass into stars, in a timescale of $\sim10-100\ $Myr, ejecting some material into a super-wind and recycling the rest of the gas into the diffuse ISM. This suppresses the bulge formation rate by  direct ``clump coalescence''  by a factor of several. However, the galactic disks do undergo significant internal evolution in the absence of mergers: clumps form and disrupt continuously and torque gas to the galactic center. The resulting evolution is qualitatively similar to bar/spiral evolution in simulations with  a more homogeneous ISM. 

\end{abstract}

\begin{keywords}
galaxies: formation --- galaxies: evolution --- galaxies: active --- 
star formation: general --- cosmology: theory
\vspace{-1.0cm}
\end{keywords}

\vspace{-1.1cm}
\section{Introduction}
\label{sec:intro}

In recent years, considerable attention has been paid to what appears to be a population of massive ($M_{\ast}\sim10^{10}-10^{11}\,\msun$), gas-rich ($M_{\rm gas}/(M_{\rm gas}+M_{\ast})\sim0.3-0.7$), rapidly star forming ($\dot{M}_{\ast}\sim10-200\,\msun\,{\rm yr^{-1}}$) ``disk'' galaxies (flattened and apparently rotationally supported, albeit with the ratio of rotational to random velocity $V/\sigma\sim\,$a few) at redshifts $z\sim2-3$ 
\citep[see e.g.][]{chapman:submm.lfs,
greve:smg.co.properties.vs.ell.models,
tacconi:smg.maximal.sb.sizes,
genzel:highz.rapid.secular,
tacconi:high.molecular.gf.highz,
shapiro:broad.halpha.emission.highz.gal,
daddi:highz.gal.high.fgas,
forsterschreiber:z2.disk.turbulence,
forster-schreiber:2011.ifu.clump.optical.obs}. 
The most extreme star-forming populations at these redshifts (with $\dot{M}_{\ast}\sim100-3000\,\msun\,{\rm yr^{-1}}$) appear to be major merger-induced starbursts, but these ``disky'' systems dominate the intermediate-luminosity population which includes a large fraction of the total SFR density  \citep[see references above and][]{tacconi:smg.mgr.lifetime.to.quiescent}. The morphology of the young stars (observed in the rest-frame UV and optical light) and star-forming or molecular gas (e.g.\ H$\alpha$) is characteristically irregular (``clumpy'' or ``clump-chain'' morphologies), with a significant fraction (tens of percent) of the light in $\sim$a few massive ($\gg10^{8}\,\msun$), $\sim$kpc-scale clumps or blobs \citep{griffiths:1994.hst.clump.gal,cowie:1995.hst.clump.gal,giavalisco:1996.hst.clump.gal,abraham:1996.hst.clump.gal,elmegreen:2004.chain.gal,elmegreen:2004.chain.gal.faceon,elmegreen:2005.hudf.morphologies.clumps,kriek:z2.hubble.sequence,genzel2011:clumps,swinbank:clumps}.

Theoretically, the origin of these morphologies is controversial. Some early models argued they were irregular because of ongoing mergers \citep{somerville:sam}; observations of local analogues (and studies in which these systems are mock-observed as if they were at high redshifts) suggest that this is true for least some  fraction of the observed systems \citep{overzier:lbgs.clumpy.disks.are.mergers,overzier:local.lbgs.w.clumps.vs.mergers,
petty:local.clumpy.systems.same.as.highz}. However, the constraints on the structure and kinematics of the massive $z \sim 2$ systems suggest that a sizeable fraction are more quiescent\footnote{By ``quiescent'' here, we mean non-interacting/non-merging, as opposed to ``quenched'' or non-star forming.} (\citealt{shapiro:highz.kinematics,forsterschreiber:z2.disk.turbulence,forster-schreiber:2011.hiz.gal.morph}), although this is not definitive evidence against mergers (\citealt{robertsonbullock:highz.disk.vs.model,
hammer:hubble.sequence.vs.mergers}). And the merger rate may not be sufficiently high to explain the abundance of these intermediate-SFR systems \citep[e.g.][]{dekel:cold.streams,stewart:merger.rates,hopkins:merger.rates,hopkins:sb.ir.lfs}.

An alternative explanation for the observed morphologies is that clumpiness stems from disk fragmentation. Any gas disk in which cooling is efficient (cooling time less than the dynamical time so that turbulent support dissipates in a crossing time, which is easily satisfied here) will fragment at the Toomre scale, $R_{T}\sim \sigma^{2}/\pi\,G\,\Sigma$ or $M_{T}\sim \sigma^{4}/\pi\,G^{2}\,\Sigma$ \citep{toomre:Q}; for disks which equilibrate at marginal stability (Toomre $Q\sim1$), this is just $R_{T}\sim f_{\rm gas}\,R_{\rm disk}$ ($M_{T}\sim f_{\rm gas}^{3}\,M_{\rm disk}$).\footnote{We have dropped numerical pre-factors that depend on the detailed disk structure. Also we could take $f_{\rm cold}$ instead of $f_{\rm gas}$ to reflect the dynamically thin/cold stellar components, although in detail the scaling for a gas plus stellar system, while qualitatively similar, becomes much more complex.} In the Milky Way, these clumps  correspond to massive GMCs ($R\sim100\,$pc, $M\sim10^{6}\,\msun$). But in a massive, gas-rich disk, the scale-length and mass of the resulting clumps can be as large as $\sim$kpc and $\sim10^{8}-10^{9}\,\msun$, respectively, similar to the structures observed. 

This instability has been well-known for $\sim40$ years. The implications of such clumping in the ISM are less clear, however. If there were no feedback from stars to expel mass from GMCs/clumps, they would (locally) collapse in a few free-fall times and turn most of their mass into stars. They would then sink to the galaxy center under the influence of dynamical friction on a timescale $t_{\rm sink}\sim t_{\rm orbit}\,M_{\rm disk}/M_{T}$; for GMCs, this would take a Hubble time, but for massive clumps in the high redshift systems of interest, this could occur in a few disk orbital periods ($\sim30-60$ clump dynamical times, or $\sim0.5\,$Gyr). This sinking would quickly build up a $\sim$kpc-scale bulge \citep{shlosman:1993.clumpy.disk.instab.sims,noguchi:1999.clumpy.disk.bulge.formation}. This is also well-known and has been seen since the first simulations of very gas-rich disks \citep{hernquist.89,barnes.hernquist.91,shlosman:1993.clumpy.disk.instab.sims,barneshernquist96,sommerlarsen99:disk.sne.fb,
robertson:cosmological.disk.formation}.

However, in the MW and nearby galaxies, typical GMCs are disrupted by stellar feedback after a few clump dynamical times (a few Myr), and recycle $\sim95\%$ of their mass back into the ISM \citep{zuckerman:1974.gmc.constraints,williams:1997.gmc.prop,evans:1999.sf.gmc.review,evans:2009.sf.efficiencies.lifetimes}. This appears to be true for both low and high mass clouds, {including} local examples with masses and sizes very similar to those observed at high redshifts \citep[$M_{\rm cloud}\sim10^{8}-10^{9}\,\msun$;][and references therein]{wilson:2003.supergiant.clumps,wilson:2006.arp220.superclumps}. 

If Toomre mass clumps were dispersed after a few dynamical times, it may be more accurate, in simulations, to treat the ISM as a ``smooth'' medium (i.e.\ not resolve the formation of individual clumps) rather than resolve clump formation but {\em not} their destruction/recycling. In fact, the one-way fragmentation seen in simulations without feedback was one of the primary motivations for modelers to include stronger prescriptions for stellar feedback in numerical galaxy models. Lacking the ability to explicitly resolve the processes that destroy clouds, this commonly amounted to assigning the gas a much larger ``effective pressure'' than its nominal thermal pressure so that runaway collapse could not occur.  Such an effective pressure was intended to mimic a {\em global} average turbulent pressure maintained by a sub-grid process of continuous clump formation and destruction \citep{hernquist.89,springel:multiphase}. This was also motivated by cosmological simulations -- the one-way break-up of gas-rich disks was  a significant factor preventing the formation of realistic disk galaxies and led to simulations which uniformly predicted  bulge-to-disk ratios significantly larger than observed \citep{sommerlarsen99:disk.sne.fb,robertson:cosmological.disk.formation,governato04:resolution.fx,governato:disk.formation}.   With the recent observations of these ``clumpy'' high-redshift galaxies, however, renewed interest has been paid to ``weak feedback'' models of the ISM \citep{immeli:fragmentation.vs.clump.properties,bournaud:disk.clumps.to.bulge,bournaud:chain.gal.model,elmegreen:classical.bulges.from.clumps,agertz:disk.fragmentation.model,dekel:2009.clumpy.disk.evolution.toymodel,ceverino:2010.clump.disks.cosmosims}.

Without an {\em a priori} resolved model of stellar feedback, the dominant feedback physics and their effects remain uncertain. \citet{genel10} considered the evolution of gas-rich disks including various phenomenological models in which stellar feedback was parameterized by a ``wind efficiency'' (in short, for every mass $\Delta M_{\ast}$ in stars formed, a mass $\eta\,\Delta M_{\ast}$ is automatically kicked out of the galaxy). They showed that for order-unity wind efficiencies of the sort needed in cosmological models to explain the shape of the galaxy mass function \citep{oppenheimer:recycled.wind.accretion}, clumps form but are disrupted by feedback before turning a large fraction of their mass into stars and so do not spiral to the center. Observations have now suggested comparable wind mass-loading factors in the clumps \citep{genzel2011:clumps,newman:z2.clump.winds}.  These models, while a powerful ``proof of concept,'' are not able to {\em predict} the wind properties or detailed ISM structure. The  studies of clumpy disks that have  included  physically-motivated models of stellar feedback have focused only on thermal heating from SNe; however, at the high gas densities of typical ``clumps'' ($\gtrsim100\,{\rm cm^{-3}}$) gas which is shock-heated by SNe ejecta will cool in a timescale $>100$ times shorter than its dynamical time, so these models unsurprisingly find similar results (in terms of clump survival) to feedback-free cases \citep[][]{elmegreen:classical.bulges.from.clumps,ceverino:2010.clump.disks.cosmosims}. And at higher densities, in fact, the SNe timescale becomes longer than the local collapse/free fall time, so they cannot act efficiently \citep{evans:2009.sf.efficiencies.lifetimes}.

A large number of feedback mechanisms may drive turbulence in the ISM and help disrupt GMCs, including: photo-ionization, stellar winds, radiation pressure from UV and IR photons, proto-stellar jets, cosmic rays, supernovae, and gravitational cascades from large scales \citep[e.g.][and references therein]{mac-low:2004.turb.sf.review}.  In \citet{hopkins:rad.pressure.sf.fb} (\paperone) and \citet{hopkins:fb.ism.prop} (\papertwo) we developed a new set of numerical models to incorporate feedback on small scales in GMCs and star-forming regions, in simulations with pc-scale resolution.\footnote{\label{foot:url}Movies of these simulations are available at \movieurl} These simulations include the momentum imparted locally (on sub-GMC scales) from stellar radiation pressure, radiation pressure on larger scales via the light that escapes star-forming regions, HII photoionization heating, as well as the heating, momentum deposition, and mass loss by SNe (Type-I and Type-II)  and stellar winds (O star and AGB).  The feedback is tied to the young stars, with the energetics and time-dependence taken directly from stellar evolution models.     Our models also include realistic cooling to temperatures $<100\,$K, and a treatment of the molecular/atomic transition in gas and its effect on star formation. We showed in Papers I \& II that these feedback mechanisms produce a quasi-steady ISM  in which giant molecular clouds form and disperse rapidly, after turning just a few percent of their mass into stars.   This leads to an ISM with phase structure, turbulent velocity dispersions, scale heights, and GMC properties (mass functions, sizes, scaling laws) in reasonable agreement with observations. In \citet{hopkins:stellar.fb.winds} (\paperthree), we showed that these same models of stellar feedback produce the elusive winds invoked in almost all galaxy formation models; the combination of multiple feedback mechanisms give rise to massive, multi-phase winds having a broad distribution of velocities, with material both stirred in local fountains and unbound from the disk. 

In this paper, we use these models to study  massive, GMC-analogue clumps in high-redshift star-forming disks, and their implications for bulge formation.

\begin{figure*}
    \centering
    \plotsidesize{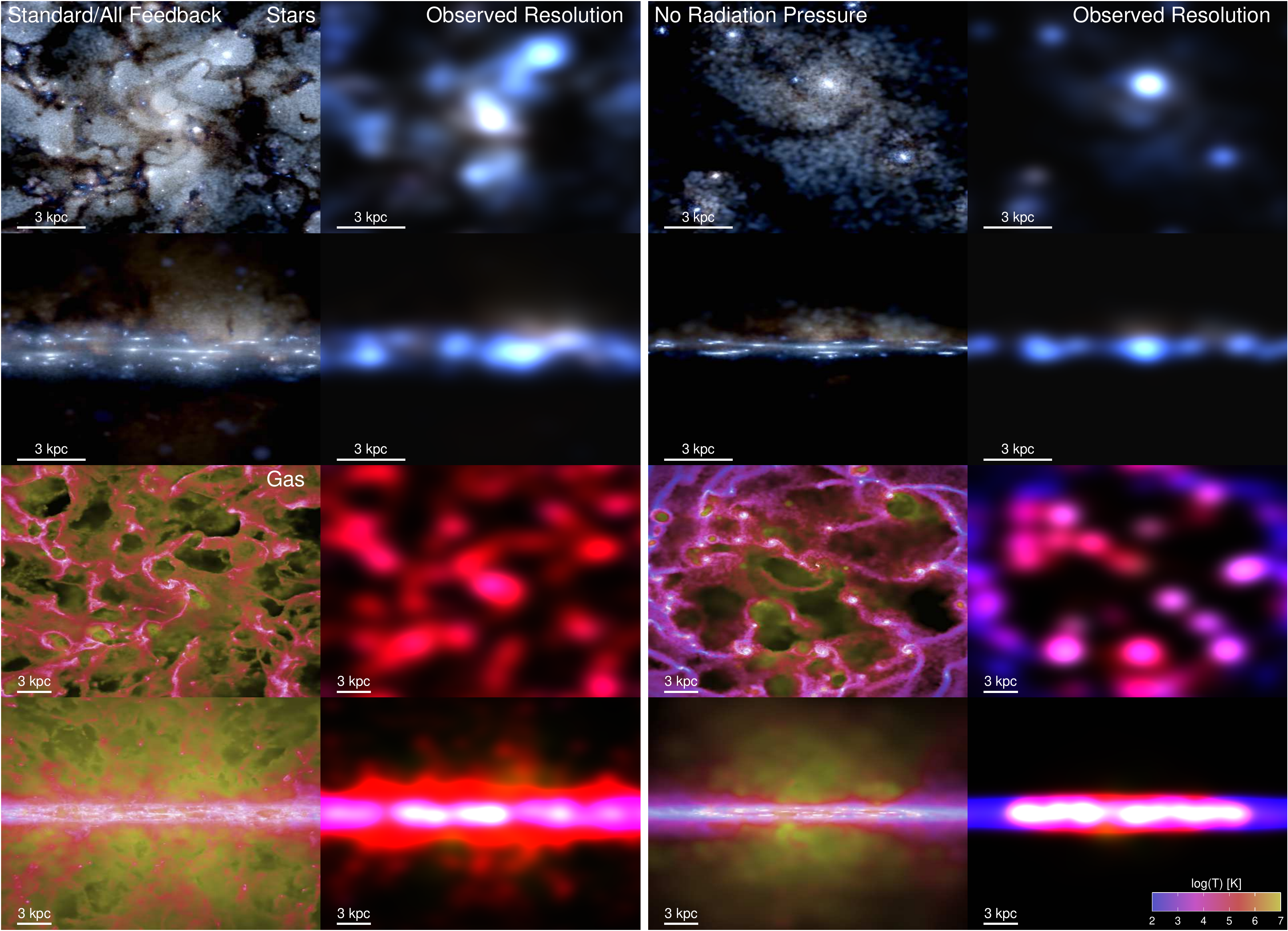}{1.0}
    \caption{Morphology of the gas and stars in a simulation of a massive $z\sim2-4$ rapidly star-forming disk with $\dot{M}_{\ast}\sim100\,\msun\,{\rm yr^{-1}}$. Images are shown
    at $\sim2$ orbital times when the disk is 
    in a feedback-regulated steady-state. We show face-on (upper) and edge-on (lower) projections.
    We compare two simulations: our ``standard'' model with all feedback mechanisms enabled ({\em left} two columns), 
    and a model with identical conditions but radiation pressure (the dominant feedback mechanism) disabled ({\em right} two columns). 
    {\em Top:} Stars. The image is a mock $ugr$ (SDSS-band) composite, 
    with the spectrum of all stars calculated from their known age and metallicity, 
    and dust extinction/reddening accounted for from the line-of-sight dust mass.$^{\ref{foot:stellarimages}}$
    For each model, the {\em left} column is an image with effectively infinite resolution and depth (brightness follows a logarithmic scale with a stretch of $\approx3\,$dex); the {\em right} column is the same image, convolved with a Gaussian PSF typical of HST WFC3 observations (FWHM $1.4\,$kpc $=0.16''$ at $z=2$) and with a linear stretch.
    {\em Bottom:} Gas. Brightness encodes projected gas density; color encodes gas temperature 
    with blue/white being $T\lesssim1000\,$K molecular gas, pink $\sim10^{4}-10^{5}$\,K 
    warm ionized gas, and yellow $\gtrsim10^{6}\,$K hot gas. 
   Again, {\em left} is at extremely high resolution/depth (logarithmically 
    scaled with a $\approx6\,$dex stretch from $\sim10^{-1}-10^{5}\,\msun\,{\rm pc^{-2}}$); {\em right} is the same with PSF convolution typical of 
    e.g.\ SINFONI AO (FWHM $1.7\,$kpc $=0.2''$ at $z=2$) and linear stretch.
    Gravitational collapse forms massive kpc-scale star cluster complexes
    that give rise to the clumpy morphology (edge on, similar to 
    ``clump-chain'' galaxies) that is further enhanced in the optical by 
    patchy extinction. The clumps have considerable sub-structure.
    With all feedback included, strong outflows are present, emerging both from the complexes and 
    the disk as a whole, driven by the massive star formation rate. These outflows maintain a 
    thick gas disk and disrupt many of the clumps, which continuously re-form. With radiation pressure disabled, the outflows are much weaker (some ``venting'' hot gas), the clumps collapse further, are more clearly rotationally supported, and leave less of a diffuse low-brightness stellar background (having coalesced more efficiently).
    \label{fig:morph}}
\end{figure*}

\begin{figure*}
    \centering
    \plotsidesize{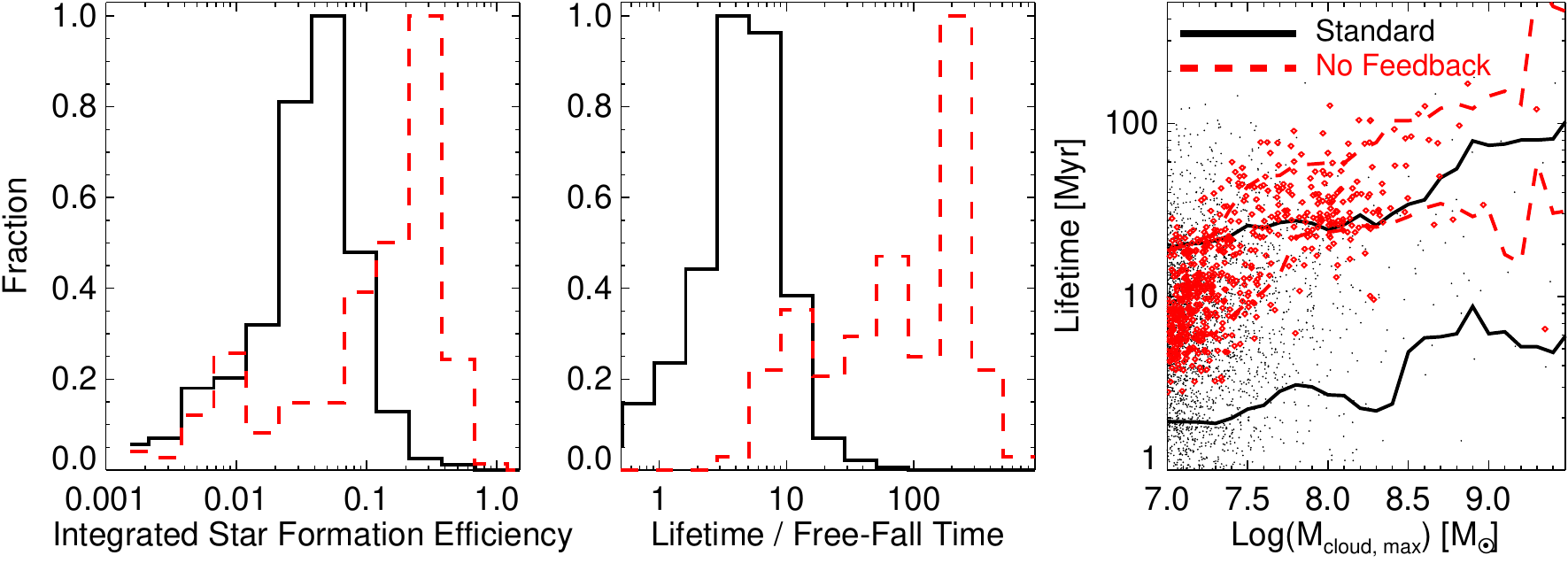}{1.0}
    \caption{Statistics of gaseous clump lifetimes and star formation efficiencies, in simulations with and without 
    stellar feedback. We identify all gas clumps that can be well-resolved ($>100$ particles) down to 
    $10\%$ of the clump mass ($>1000$ particles at peak); lifetime is defined 
    as the time above $10\%$ the maximum cloud gas mass. 
    {\em Left:} Integrated star formation efficiency (stellar mass formed in clump 
    over maximum cloud gas mass). We compare the ``standard'' model (all feedback included) to a model 
    with no feedback.
    {\em Middle:} Cloud lifetime in units of the (mass-weighted average) free-fall time $t_{\rm ff}\approx0.54/\sqrt{G\,\rho}$ within the clump. 
    {\em Right:} Lifetime versus maximum cloud gas mass. We show the result for individual clouds (small black 
    and large red points for the standard/no feedback models, respectively) and the lines show 
    the $\pm1\,\sigma$ range at each cloud mass for each model.
     With feedback present, even the most massive gas 
    clouds live for a few dynamical times and turn $\sim10\%$ of their mass into stars. 
    Without feedback, clouds persist until most of their dense gas 
    is turned into stars, giving star formation efficiencies of $\sim50\%$; their lifetimes are longer and they 
    ``end'' only because the gas is exhausted (the relic stellar clumps are much longer-lived). 
    \label{fig:lifetimes}}
\end{figure*}

\begin{figure}
    \centering
    \plotonesize{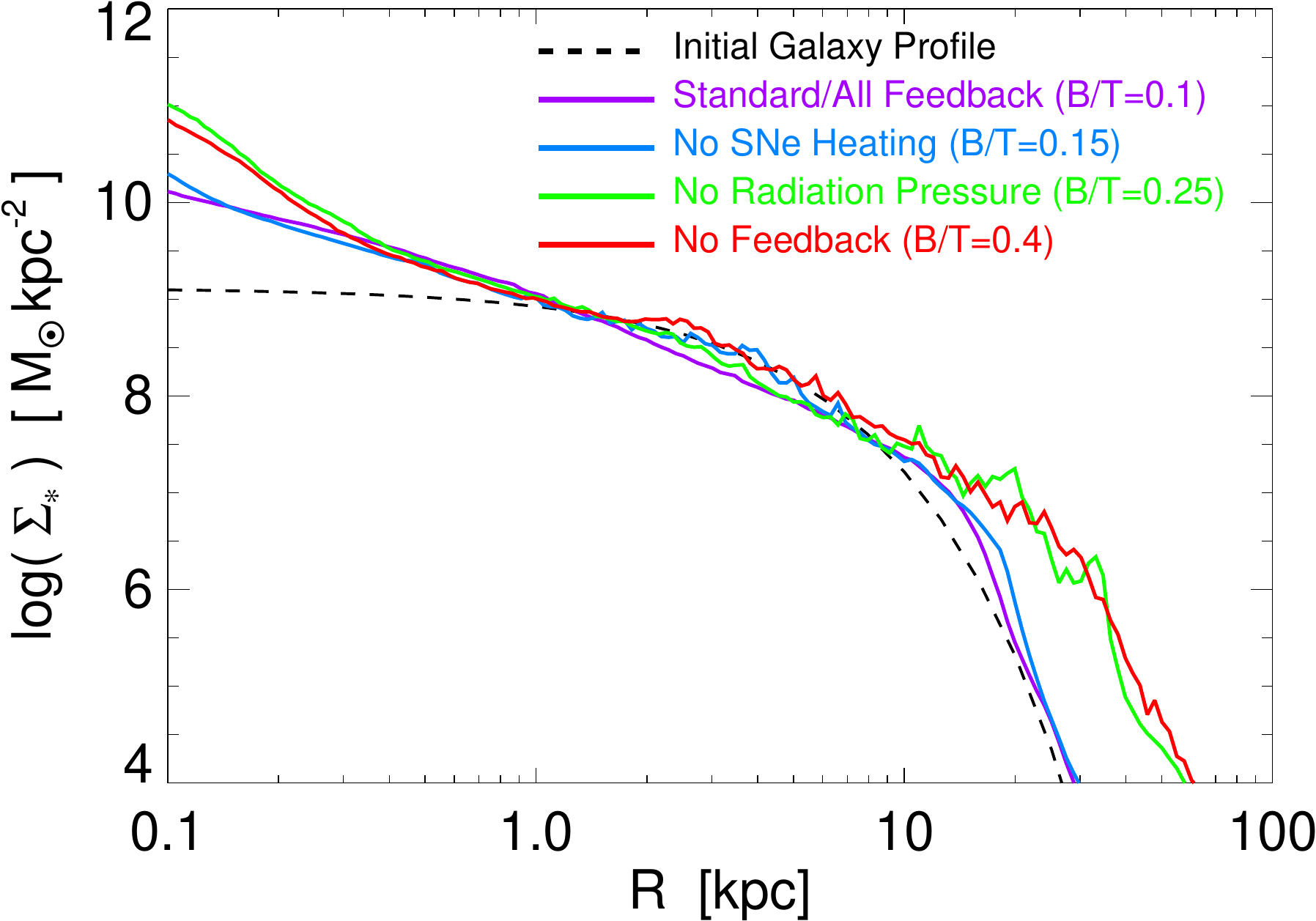}{1.0}
    \caption{Remnant stellar mass profiles for our simulations, with different stellar feedback 
    mechanisms enabled or disabled (as labeled). The profiles are averaged over $\sim100$ projections.
    The initial disk profile is shown for comparison. There is internal (rapid ``secular'') evolution in all cases: gas that loses angular momentum builds a pseudo-bulge while some gas gains angular momentum, making the outer disk larger. 
    With our standard feedback model, the bulge is small ($B/T\sim0.1$) and disky ($n_{s}\sim1$), and the outer disk does not evolve significantly. Removing SNe heating makes little difference. Removing radiation pressure, however, is nearly identical to removing all feedback, which yields a much larger, more concentrated ($n_{s}\sim2$) bulge as well as a stellar envelope at large radii that includes $\sim60-70\%$ of the mass.
    \label{fig:massprofile}}
\end{figure}

\vspace{-0.5cm}
\section{The Simulations}
\label{sec:sims}

The simulations used here are described in detail in 
\paperone\ (Sec.~2 \&\ Tables~1-3) and \papertwo\ (Sec.~2).
We briefly summarize the most important properties here. 
The simulations were performed with the parallel TreeSPH code {\small 
GADGET-3} \citep{springel:gadget}. They include stars, dark matter, and gas, 
with cooling, star formation, and stellar feedback. 

Gas follows an atomic cooling curve with additional fine-structure 
cooling to $<100\,$K, with no ``cooling floor'' imposed. 
Star formation is allowed only in dense regions above $n>1000\,{\rm cm^{-3}}$, 
at a rate $\dot{\rho}_{\ast}=\epsilon\,\rho_{\rm mol}/t_{\rm ff}$ 
where $t_{\rm ff}$ is the free-fall time, $\rho_{\rm mol}=f_{\rm H_{2}}\,\rho$ 
is the molecular gas density, and $\epsilon=1.5\%$ 
is the observed efficiency at these densities 
\citep[][]{krumholz:sf.eff.in.clouds}. We follow \citet{krumholz:2011.molecular.prescription} 
to calculate the molecular fraction $f_{\rm H_{2}}$ in dense gas as a function 
of local column density and metallicity. 
In \paperone\ and \papertwo\ we show that the galaxy structure and SFR 
are basically independent of the small-scale SF law, density threshold (provided it is high), 
and treatment of molecular chemistry.

We consider an initial galaxy model intended to represent an (intentionally extreme) high-redshift massive ``starburst disk'' (typical of high-luminosity BzK galaxies or low-luminosity SMGs -- intermediate SFR systems -- at $z\sim2-4$). The disk has a small ($B/T<0.05$) initial \citet{hernquist:profile} profile bulge ($M_{b}=5\times10^{9}\,\msun$ (chosen to be small to avoid suppressing disk instabilities and clearly isolate new bulge growth), scale-length $a=1.2\,{\rm kpc}$), an exponential stellar disk ($M_{\rm d}=3\times10^{10}\,\msun$; $r_{d}=1.6\,{\rm kpc}$ with sech$^{2}$ profile scale-height $h=130\,$pc) a gas disk ($M_{\rm g}=7\times10^{10}\,\msun$; $r_{g}=3.2\,{\rm kpc}$), 
and a dark matter halo ($M_{\rm halo}=1.4\times10^{12}\,\msun$, concentration $c=3.5$, scaled to lie on the mass-concentration relation for $z=2$ halos).   The model has $\approx10^{8}$ total particles and $5\,$pc gravitational softening.

Stellar feedback is included, via a variety of mechanisms.

(1) {\bf Local Momentum Flux} from Radiation Pressure, 
Supernovae, \&\ Stellar Winds: Gas within a GMC (identified 
with an on-the-fly friends-of-friends algorithm) receives a direct 
momentum flux from the stars in that cluster/clump. 
The momentum flux is $\dot{P}=\dot{P}_{\rm SNe}+\dot{P}_{\rm w}+\dot{P}_{\rm rad}$, 
where the separate terms represent the direct momentum flux of 
SNe ejecta, stellar winds, and radiation pressure. 
The first two are directly tabulated for a single stellar population as a function of age 
and metallicity $Z$ and the flux is directed away from the stellar center. 
Because this is interior to clouds, the systems are always optically thick, so the 
latter is approximately $\dot{P}_{\rm rad}\approx (1+\tau_{\rm IR})\,L_{\rm incident}/c$, 
where $1+\tau_{\rm IR} = 1+\Sigma_{\rm gas}\,\kappa_{\rm IR}$ accounts 
for the absorption of the initial UV/optical flux and multiple scatterings of the 
IR flux if the region is optically thick in the IR (with $\Sigma_{\rm gas}$ calculated 
for each particle given its location in the clump). 

(2) {\bf Supernova Shock-Heating}: Gas shocked by 
supernovae can be heated to high temperatures. 
We tabulate the SNe Type-I and Type-II rates from 
\citet{mannucci:2006.snIa.rates} and STARBURST99, respectively, as a function of age and 
metallicity for all star particles and stochastically determine at 
each timestep if a SNe occurs. If so, the appropriate mechanical luminosity is 
injected as thermal energy in the gas within a smoothing length (nearest 32 gas neighbors) of the star particle. 

(3) {\bf Gas Recycling and Shock-Heating in Stellar Winds:} Gas mass is returned 
to the ISM from stellar evolution, at a rate tabulated from SNe and stellar mass 
loss (integrated fraction $\approx0.3$). The SNe heating is described above. Similarly, stellar winds 
are assumed to shock locally and inject the appropriate tabulated mechanical 
luminosity $L(t,\,Z)$ as a function of age and metallicity into the gas within a smoothing length. 

(4) {\bf Photo-Heating of HII Regions}: We also tabulate the rate of production of ionizing photons for 
each star particle; moving radially outwards from the star, we then ionize each neutral gas particle (using 
its density and state to determine the necessary photon number) 
until the photon budget is exhausted. Ionized gas is maintained at a minimum $\sim10^{4}\,$K until 
it falls outside an HII region.

(5) {\bf Long-Range Radiation Pressure:} Photons which escape the local GMC (not 
absorbed in mechanism (1) above) can be absorbed at larger radii. Knowing the intrinsic SED of each star 
particle, we attenuate integrating the local gas density and gradients to convergence. 
The resulting ``escaped'' SED gives a flux that propagates to large distances, and 
can be treated in the same manner as the gravity tree to give the local net incident flux 
on a gas particle. The local absorption is then calculated integrating over a frequency-dependent 
opacity that scales with metallicity, and the radiation pressure force is imparted (and luminosity removed). 

In implementing (1)-(5), all energy, mass, and momentum-injection rates are taken  from  stellar 
population models  \citep{starburst99}, assuming a \citet{kroupa:imf} IMF, without any free parameters.
More details, numerical tests, and resolution studies (up to $10^{9}$ particles with $3.5\,$pc softening lengths) for these models are discussed in \papertwo. 

We note that some recent studies of low-resolution cosmological simulations comparing {\small GADGET} and the moving mesh code {\small AREPO} \citep{springel:arepo} have highlighted some differences between smoothed particle hydrodynamics and grid methods for some cosmological inflow problems \citep{vogelsberger:2011.arepo.vs.gadget.cosmo,keres:2011.arepo.gadget.disk.angmom,torrey:2011.arepo.disks}. However, we have also performed idealized simulation comparisons between the individual, high-resolution galaxy models here (as well as galaxy mergers) and found excellent agreement for e.g.\ the star-formation rates, gas inflow through the disk, and surface density profiles; these comparisons will be presented in \citet{hayward:arepo.gadget.mergers}. We have also re-run the simulations here with an alternative density-independent formulation of SPH (which resolves these discrepancies) as proposed in \citet{saitoh:2012.dens.indep.sph}; again we find excellent agreement. This owes in part to the resolution, as well as to the fact that the code differences are minimized when flows are super-sonic, always true here \citep[see][]{bauer:2011.sph.vs.arepo.shocks}. To the extent that there are differences, the moving-mesh methods may tend to slightly shorter clump lifetimes because stripping and mixing of the gas as it moves are more efficient (and artificial shock dissipation is reduced) \citep[see][]{sijacki:2011.gadget.arepo.hydro.tests}. 

\vspace{-0.5cm}
\section{Results}
\label{sec:morph}

Fig.~\ref{fig:morph} compares the gas and stellar\footnote{\label{foot:stellarimages}The stellar luminosity in each band is calculated from each star particle 
according to the STARBURST99 model given its age, mass, and metallicity 
(and smoothed over the appropriate kernel). We then attenuate the stars 
following the method of \citet{hopkins:lifetimes.letter}: we calculate the total dust 
column (from the simulated gas) along the line-of-sight to each star particle 
for the chosen viewing angle (assuming a constant 
dust-to-metals ratio, i.e.\, dust-to-gas equal to the MW value 
times $Z/Z_{\sun}$), and apply a MW-like extinction and 
reddening curve \citep[as tabulated in][]{pei92:reddening.curves}.} morphologies of runs with varying feedback properties. In all cases, the disks form massive ($>10^{8}\,\msun$) $\sim$kpc-scale clumps, which in turn form massive star cluster complexes. The gas densities are sufficiently high that extinction can lead to a clumpy optical morphology as well (by entirely extincting out regions in the galaxy). The clumpiness is prominent in the gas  and stars (although, with feedback present, there is a low surface brightness smooth stellar disk that is not visible at the present observational depth). Seen edge-on, the gas can resemble a ``clump-chain'' morphology. With feedback present, strong outflows arise from throughout the disk, driven by a massive star formation rate of $>100\,\msun\,{\rm yr^{-1}}$. 

In previous papers, we present detailed comparisons between our simulated disk properties and observations.  The model disk masses, rotation curves, scale lengths, and gas fractions (see \S~\ref{sec:sims}) are chosen to match those observed.   Their resulting SFRs ($50-300\,\msun\,{\rm yr^{-1}}$)  are consistent with the Kennicutt-Schmidt relation (\paperone; Fig.~11 and \papertwo; Fig.~7).   Moreover, the simulated galaxies self-regulate at Toomre $Q\approx1$, with typical line-of-sight gas velocity dispersions $\sigma\approx50-80\,{\rm km\,s^{-1}}$ and star-forming gas disk scale heights $h\approx1\,{\rm kpc}$ or $h/R\approx0.2$ (\papertwo; Fig.~9). However, 
we show in \papertwo\ that our model disks always reach $Q\sim1$ and so -- for fixed global disk properties -- $\sigma$ is almost completely independent of the SFR density and the microphysics of feedback.  The maximum clump masses are $\sim10^{9}\,\msun$ with a few clumps per disk within a factor of a few of this mass (\papertwo; Fig.~13); the corresponding clump sizes are $\sim0.3-2\,$kpc (\papertwo; Fig.~15) giving typical clump surface densities $\sim 300-1000\,\msun\,{\rm pc^{-2}}$ (\papertwo; Fig.~14). The clumps are marginally bound (and not rotationally supported, although mildly anisotropic), giving them internal velocity dispersions nearly equal to that of the ``background'' disk gas (\papertwo; Fig.~16; see also \citealt{hopkins:excursion.ism}). The winds driven off of the disk and clumps have a broad velocity distribution from $\sim200-750\ {\rm km\,s^{-1}}$ (\paperthree; Fig.~4), with globally averaged outflow rates in the high-speed superwind of $\dot{M}_{\rm wind}\approx 0.5-2\,\dot{M}_{\ast}$ (\paperthree; Figs.~6-7). All of these properties agree reasonably well with those observed \citep[see][and references therein]{genzel2011:clumps}. As shown in \papertwo\ and \paperthree, the simulations quickly establish these properties in the first couple dynamical times and maintain them in quasi-steady state as long as they are evolved, until gas is exhausted.


We now quantify the properties of the clumps in more detail and discuss their implications for the internal (``secular,'' but rapid) evolution of the galaxy.   We identify gas clumps following \papertwo; specifically, we apply the sub-halo finder SUBFIND to the gas, which employs a friends-of-friends linking algorithm with an iterative procedure to robustly identify overdensities \citep[for details and tests, see][]{springel:cluster.subhalos}. In \papertwo\ we show that visual inspection and calculation of the binding energy of clumps confirm that this correctly identifies the obvious clumps; changing linking lengths and other numerical quantities has a weak effect on the properties here. Because the gas particles are unique, we can link clumps in time between different narrowly spaced snapshots ($\Delta t\approx10^{6}\,$yr) to construct a clump ``merger tree.''\footnote{Here the ``main branch'' is defined by the most massive clump in the tree at each time. Mergers of other clumps are generally small -- most growth is via accretion of ``non-clump'' gas, so we neglect the ``merged branches.''} We then obtain the growth/decay in mass with time for each clump; we define the clump ``lifetime''  as the time spent above $10\%$ of the maximum clump mass (this cut is arbitrary, but not critical to our results). We can also define the total clump star formation efficiency by integrating the SFR of all clump-associated particles at each time.

Fig.~\ref{fig:lifetimes} plots the distribution of gas clump lifetimes and star formation efficiencies. 
As shown in \papertwo, the most massive gas clumps do tend to be longer-lived; however, with feedback present their lifetimes are still just a few internal dynamical times ($\sim10-100\,$Myr; this is consistent with the -- highly uncertain -- observed stellar populations in clouds in e.g.\ \citealt{elmegreen:2009.clump.properties.hudf,forster-schreiber:2011.ifu.clump.optical.obs}, but more detailed observations would present a useful constraint on the predictions here).  This is short compared to their inspiral time by dynamical friction. About $\sim5-20\%$ of the peak gas mass of the clumps is turned into stars; this is a factor of a few higher than the star formation efficiency of local GMCs \citep[as expected by theoretical arguments; see][]{murray:molcloud.disrupt.by.rad.pressure}, but much less than the order-unity fraction needed to conserve the ``relic'' stellar clump as a bound entity.\footnote{Many of the lowest mass-clouds, in both the models with and without feedback, have short lifetimes. The reason is, as shown in \papertwo, many of these are never bound but are simply transient overdensities. We also caution that they may still be subject to resolution effects. But in any case these represent a small fraction of the total gas mass and SFR.}

Without feedback, the gas collapses to higher densities rapidly, so the free-fall time $\propto \rho^{-1/2}$ becomes very short and the gas lifetime in units of that free-fall time becomes as long as $\sim10-100\,t_{\rm ff}$. This runaway collapse leads to large star formation efficiencies $\sim20-100\%$, with a mean of $\sim50\%$ (it is not $100\%$ because there is always some low-density material associated with clouds). Note that the lifetime of the ``gas clump,'' defined as we consider it, may then still be relatively short, $\sim20-100\,$Myr; this is because most of the mass has turned into stars,  a large fraction of which remains self-gravitating and will survive in a bound stellar clump until reaching the galaxy center. 

Fig.~\ref{fig:massprofile} shows the resulting mass profiles at $t=500\,$Myr (for the initial disk model, $t_{\rm dyn}\equiv R_{e}/V_{c}(R_{e})\approx12\,$Myr, so this is 40 (7) dynamical (orbital) times at $R_{e}$). By this point the gas is largely exhausted or blown out in all models and the evolution has slowed down; most ``sinking'' clumps have coalesced. That is not to say this is a ``final state'' -- the system undergoes continuous long-term collisionless stellar/halo internal and secular evolution, and in a cosmological context it would evolve significantly over $\sim$Gyr timescales. We are simply comparing the {relative} effects of including or not including feedback. 

In all the simulations, there is some internal/``secular'' evolution away from the initial disk conditions. The gas is still asymmetric and turbulent, hence there are gravitational torques \citep[see e.g.][]{gammie:2001.cooling.in.keplerian.disks,hopkins:inflow.analytics,dekel:2009.clumpy.disk.evolution.toymodel,bournaud:2011.agn.fueling.by.clumps,forbes:2011.thick.disk.torque.evol}. Some mass is channeled to the central $\sim$kpc, building up the ``bulge''; the angular momentum loss required for this is absorbed by the outer disk (basically the equivalent of an outer Lindblad resonance), which then migrates outwards into a more extended, flattened disk ``envelope'' \citep[which may, in some cases, be directly observed;][]{elmegreen:2006.rings.as.linblad.for.clumps}. 

With feedback, this evolution is modest; fitting a bulge-disk decomposition to the profile (with a Sersic-profile bulge and exponential disk), we obtain ($B/T,\ n_{s},\ R_{d})=(0.10,\,1.1,\,4.3\,{\rm kpc})$ for the bulge mass fraction, Sersic index (lower $n_{s}<2$ representing more disk-like ``pseudobulges''), and disk effective radius (compare to the initial $R_{d}=2.7$\,kpc).\footnote{We caution that the best-fit bulge profile (both with and without feedback) is quite sensitive to the exact fitting procedure, since these galaxies (being clumpy and asymmetric) do not have a well-defined ``center'' at the sub-kpc level (so it makes a large difference whether we center on a clump density maximum or between clumps). Here we use an iterative elliptical aperture density smoothing to isolate the central density maximum \citep[see][]{hopkins:cusps.ell}, which gives the maximum ``bulge'' component of any fitting methods we explore. Our qualitative conclusions on relative bulge formation efficiencies are the same with each method. The outwards evolution of the disk, however, is a much more robust indication of the degree of internal evolution and is completely insensitive to these fitting details.}
With no feedback, we see a dramatic central rise in the profile indicative of a large bulge, and a much more pronounced outwards evolution of the disk; ($B/T,\ n_{s},\ R_{d})=(0.37,\,1.9,\,10\,{\rm kpc})$. Fig.~\ref{fig:massprofile} compares several intermediate cases as well:   if we remove radiation pressure, including only sources of gas ``heating'' (SNe, stellar wind shock-heating, and HII heating), the evolution is only slightly less severe than the feedback-free case. But if we include only radiation pressure, the profile is only somewhat more evolved than the standard feedback case. If, however, we remove long-range radiation pressure forces, keeping only local radiation pressure in regions optically thick in the IR, then it is again more similar to the feedback-free case. As we show in \paperone\ and \paperthree, the long-range photon momentum is critical for driving winds; the short-range pressure tends to drive local turbulence within clumps, but may not completely disperse them. 

\vspace{-0.5cm}
\section{Discussion}
\label{sec:discussion}

We have simulated the isolated internal (rapid secular) evolution of gas-rich (initial $f_{\rm gas}\approx0.6$), massive ($10^{11}\,\msun$), disk-dominated galaxies, designed to resemble the most  massive star-forming disks observed at redshifts $z\sim2-3$.   Our simulations  include explicit, physical models for stellar feedback from a variety of sources:   radiation pressure (from UV and IR photons), HII photoionization, SNe, and stellar winds. In \papertwo, we show that these models agree reasonably well with observed systems in a wide range of measurable properties: effective radius, mass, gas fraction, and rotation curve (by construction), as well as gas velocity dispersion, SFR, Toomre $Q$ parameter, gas disk scale height ($h/R\approx0.2$), and characteristic clump masses and sizes. Here we have quantified the galaxy morphology and the evolution of its stellar mass profile with different feedback mechanisms included or removed from the simulations. 

We find that the models all experience some degree of internal evolution. This is expected: they are all gas-rich, disk-dominated systems in which the cooling time is short relative to the dynamical time, so are both locally and globally gravitationally unstable. We showed in \papertwo\ that, independent of the details of feedback, the disks converge to $Q\sim1$, which means that the gas velocity dispersion in all cases is essentially identical.\footnote{This does not mean the dispersion cannot be ``driven by'' feedback; it simply means that once systems reach $Q>1$, excess energy is lost (e.g.\ in winds) or dissipated, without new star formation, and once $Q<1$, they collapse until sufficient stars form (or until sufficient gas is exhausted) to bring $Q$ back up to unity.} This means that the characteristic Toomre length and mass scales are set  by global properties and are similar in all of our simulations, even with very different feedback physics: $M_{\rm Toomre}\sim h^{2}\,\Sigma_{\rm gas}\sim f_{\rm gas}^{3}\,M_{\rm disk}\sim 10^{9}\,\msun$. The size and mass scales for the most massive clumps are therefore similar in all our simulations. In \papertwo\ we explicitly show that the mass function of gas clumps is also broadly similar.
These clumps, then, are the generic extension of GMCs in the MW (which has a lower Toomre mass) to more gas-rich massive systems \citep[although with some subtle distinctions; see][]{krumholz:2012.universal.sf.efficiency}. 

However, the effects of the resulting clumps on bulge formation depends strongly on the inclusion of feedback. With no feedback, the clumps collapse rapidly and turn $\sim50\%$ of their mass into stars. They have no alternative but to sink to the center in a few orbital periods: this gives rise to a massive, $\sim$kpc-scale bulge, and causes substantial migration of the remaining disk gas and stars to larger radii (since the angular momentum of the sinking clumps must be taken up by the remaining gas/stars, and tends to do so near the equivalent of an outer Lindblad resonance).   These effects are, broadly speaking,  well-known results of strong internal/secular evolution, though some of the details differ in this very clumpy limit (and may differ further with ongoing accretion).   For an isolated disk, this evolution  is self-terminating once sufficient bulge mass is formed, since the outward disk migration and inner bulge formation both stabilize the system; this generically leaves a large low-density disk ``envelope'' rather than a pure bulge \citep[a well-known result shown in][]{elmegreen:2005.exponential.disk.by.clump.form.hiz.small,bournaud:disk.clumps.to.bulge}.

In our standard feedback model -- with all mechanisms included --   even the most massive clumps tend to be disrupted after they turn $\sim5-20\%$ of their mass into stars. Even if those stars can remain bound after the clump is dispersed (in which case they will still tend to sink to the center, albeit on a factor $\sim10$ longer timescale) this suppresses the efficiency of bulge formation via clump coalescence by an order of magnitude. And indeed we see much less dramatic signatures of internal/secular evolution: both a smaller bulge, and much weaker evolution of the outer disk.  The resulting evolution is  less rapid compared to one-way fragmentation, and produces a more traditional pseudobulge.   However, the clump morphologies are not radically different; if anything, the clumps appear somewhat {\em more} prominent with feedback, because they are prevented from internally collapsing to  small sizes. They are also less rotationally supported in better agreement with the observations \citep{forster-schreiber:2011.ifu.clump.optical.obs,genzel2011:clumps,swinbank:clumps}\footnote{The more prominent rotation -- since there is no vertical feedback support -- in the no-feedback case leads to the mini-spiral structure seen in some cases in Fig.~\ref{fig:morph}; for a more detailed discussion of the rotation support in the SNe-only case, we refer to \citet{ceverino:2012.clump.rotation}.}

A natural prediction of this model is that the old stars should be more smoothly distributed than the gas and young stars (somewhat evident in Fig.~\ref{fig:morph}).\footnote{This is always true to some extent, however it will plainly be moreso if the clumps themselves are short-lived and transient, rather than long-lived features.} 
A related prediction is plainly that the clumps are relatively short-lived, with gas-phase lifetimes shown in Fig.~\ref{fig:lifetimes}; to some extent this should be manifest as young stellar population ages for the stars physically bound to the clumps. 
Testing these predictions requires observations that can probe the rest-frame near infrared structure of these galaxies. Moreover, patchy obscuration (given the very large gas densities) can make even the oldest background populations appear clumpy \citep[see][]{genel10}. And as also noted in \citet{genel10}, the fact that clumps form from {\em both} gas and stars means that -- especially if the star formation efficiency is low and lifetimes are short -- there will always be contributions to from pre-existing disk stars (with $\sim$Gyr ages), which will introduce contaminating old stellar populations to all clumps as well as gradients in the clump ages with galactocentric radius (the clump age gradients simply tracing the pre-existing disk gradients). We  therefore defer more detailed predictions of clump appearance in different observed wavelengths to future work, in which we will use radiative transfer models to construct mock observations. However, preliminary observational studies have already shown that clumps dominate the light only in young stellar populations and represent transient local fluctuations in luminosity; when constructing stellar mass maps, clumps become far less significant \citep{wuyts:2012.no.clumps.in.candels.stellarmass}. A rigorous comparison of the simulations here (using mock observations in the relevant observed bands) and these observations will be the subject of future work, but preliminary comparisons suggest that they strongly disfavor the ``weak/no feedback'' models (Wuyts et al., in prep). 

Which feedback mechanisms are most important in dissociating clumps and recycling gas into the ISM? We have shown in \papertwo\ and \paperthree\ that in these dense, gas-rich systems radiation pressure is dominant over sources of thermal heating such as shocked SNe ejecta, stellar winds, and HII photoionization. This is because the gas densities within clumps are so large that the cooling time is significantly shorter than the dynamical time.   
This explains why previous simulations with feedback included only in the form of thermal heating from SNe have seen very little effect on the evolution of  massive star-forming disks (see \S~\ref{sec:intro}).

Within the context of radiation pressure, we also find that it is specifically the ability to drive winds out of clumps (and, possibly but not necessarily, the galaxy) that has the largest effect on long-term clump survival and bulge formation. This supports the conclusions of \citet{genel10}, who found the same in fully cosmological simulations, but with a phenomenological wind model (see \S~\ref{sec:intro}). Since the models here self-consistently produce winds with comparable global mass-loading factors (see \paperthree), many of the conclusions should be the same (and we should expect to find similar results in cosmological simulations). In detail, the global wind mass-loading factors here are relatively modest, $\dot{M}_{\rm wind}\approx\dot{M}_{\ast}$ (Fig.~6-7 in \paperthree), a factor of $\approx4$ lower than those in \citet{genel10} and closer to the values (at this mass and redshift) typically invoked in cosmological models to reproduce the galaxy stellar mass function \citep{oppenheimer:recycled.wind.accretion}.\footnote{For a detailed comparison of the wind mass-loading, structure, velocities, and phases with observations and cosmological constraints, we refer to \paperthree.} But the implications for clump survival are similar. The reason is that the feedback models here give rise to large {\em local} outflows and fountains (which do not appear in the \citealt{genel10} model), driving turbulence and outflows from individual clumps at modest velocities (several tens of ${\rm km\,s^{-1}}$). This is sufficient to unbind gas from a clump but not from the galaxy as a whole. Only a fraction of the material in these fountains is then further accelerated into the galactic outflow at $>100\,{\rm km\,s^{-1}}$. A generic expectation of our more realistic models is that there should be significantly more mass at modest velocities compared to the true ``super-wind;'' therefore even modest mass-loading factors of $\gtrsim0.5\,\dot{M}_{\ast}$ at high velocities are consistent with our predictions here and probably correspond to sufficient local outflows and fountains to suppress clump inspiral.

\citet{krumholz:2010.highz.clump.survival} used analytic  models to consider whether or not radiation pressure would disrupt the clumps observed in high redshift galaxies, and concluded that it would not. There are a number of important differences between the full non-linear results here and their simplified model. The clouds here are clumpy and marginally bound, with a large fraction of the clump mass only transiently associated with a given cloud (as opposed to tightly bound or virialized; see \papertwo).  A combination of disk shear, gravitational instability and feedback operate simultaneously (lowering the threshold for efficient feedback). \citet{krumholz:2010.highz.clump.survival} assumed a minimal momentum flux from young stars (about the flux from the single scattering limit); considering the additional contributions from SNe and stellar wind momentum flux, and multiple scatterings in the IR, the momentum flux here is a factor of several larger. And they assumed a star formation efficiency of $\sim1\%$ per clump dynamical time, even though they argue that this fails to prevent collapse. In \paperone\ and \papertwo, we show that whenever feedback is insufficient to disperse a GMC, runaway local collapse proceeds until the SFR increases to the level needed to once again further suppress collapse \citep[see also][]{tasker:2011.photoion.heating.gmc.evol,dobbs:2011.why.gmcs.unbound}. Thus a higher efficiency {always} arises when feedback is ``initially'' insufficient. Overall, our calculations demonstrate that an instantaneous star formation efficiency of just $\sim1-5\%$ (or an integral efficiency of $\sim5-20\%$)  is more than sufficient to unbind (most of) the clumps.

Given the stellar feedback mechanisms above, the models here predict local outflows from clumps in qualitative agreement with observations \citep{genzel2011:clumps,newman:z2.clump.winds}. Both our models here and those in \citet{genel10} indicate that, if the wind mass is comparable to that observed, clumps will, in fact, tend to be disrupted.  This is also consistent with what is observed at low redshifts (albeit in different galaxy types):  in some starburst and mergers the densities and velocity dispersions are similar to these high-redshift systems, and so the characteristic Toomre mass is similar; indeed, $\sim100$ GMCs with masses $\sim10^{8}-3\times10^{9}\,\msun$ and radii $\sim0.1-1\,$kpc have been observed in the local Universe in such extreme systems \citep[see][and references therein]{rand:1990.m51.superclumps,planesas:1991.1068.superclumps,wilson:2003.supergiant.clumps,wilson:2006.arp220.superclumps}. A combination of constraints from stellar population ages and masses, the maximum mass of star clusters observed in the same galaxies, the specific frequencies of bound/globular clusters, and the cluster formation rates all suggest that these clumps have (gas) lifetimes $\lesssim100\,$Myr, turn only a modest fraction $\sim10\%$ of their mass into stars, and typically do not remain bound. In future work, we will extend our low-redshift models to starbursts and mergers to compare directly with these observations, but preliminary comparison suggests that they self-regulate in similar fashion.


Our calculations do not include the  inflow of gas via continuous accretion and mergers in a cosmological context. This allows us to robustly distinguish the effects of true isolated evolution and disk instability from e.g.\ ongoing minor mergers. But as a result, our conclusions should be viewed as statements about the relative efficiency of internal evolution with and without feedback, not statements about absolute bulge masses/structure. If the models here are representative of evolution over a few hundred Myr, then a minor merger of mass ratio 10:1 or larger would be able to induce comparable bulge growth, and a major merger would change the system's properties by more than several Gyr of evolution in this mode. Continuous accretion may, however, accelerate the evolution produced by the short-lived clumps and/or effect angular momentum transfer \citep[see][]{keres:hot.halos}. In future work, we will extend these models to cosmological zoom-in simulations to explore these questions in more detail.

\vspace{-0.7cm}
\acknowledgments 
We thank our referee, Avishai Dekel, for a number of insightful comments. We also thank Stijn Wuyts, Reinhard Genzel, Sarah Newman, and Sandy Faber for helpful discussions throughout the development of this manuscript.
Support for PFH was provided by NASA through Einstein Postdoctoral Fellowship Award Number PF1-120083 issued by the Chandra X-ray Observatory Center, which is operated by the Smithsonian Astrophysical Observatory for and on behalf of NASA under contract NAS8-03060. EQ is supported in part by the David and Lucile Packard Foundation.  NM is supported in part by NSERC and by the Canada Research Chairs program. DK acknowledges support from NASA through Hubble Fellowship grant HSTHF-51276.01-A. The simulations in this paper were run on the Odyssey cluster supported by the FAS Science Division Research Computing Group at Harvard University. 
\\

\vspace{-0.1cm}
\bibliography{/Users/phopkins/Documents/lars_galaxies/papers/ms}

\begin{thebibliography}{102}
\expandafter\ifx\csname natexlab\endcsname\relax\def\natexlab#1{#1}\fi

\bibitem[{{Abraham} {et~al.}(1996){Abraham}, {Tanvir}, {Santiago}, {Ellis},
  {Glazebrook}, \& {van den Bergh}}]{abraham:1996.hst.clump.gal}
{Abraham}, R.~G., {Tanvir}, N.~R., {Santiago}, B.~X., {Ellis}, R.~S.,
  {Glazebrook}, K., \& {van den Bergh}, S. 1996, \mnras, 279, L47

\bibitem[{{Agertz} {et~al.}(2009){Agertz}, {Teyssier}, \&
  {Moore}}]{agertz:disk.fragmentation.model}
{Agertz}, O., {Teyssier}, R., \& {Moore}, B. 2009, \mnras, 397, L64

\bibitem[{{Barnes} \& {Hernquist}(1996)}]{barneshernquist96}
{Barnes}, J.~E., \& {Hernquist}, L. 1996, \apj, 471, 115

\bibitem[{{Barnes} \& {Hernquist}(1991)}]{barnes.hernquist.91}
{Barnes}, J.~E., \& {Hernquist}, L.~E. 1991, \apjl, 370, L65

\bibitem[{{Bauer} \& {Springel}(2012)}]{bauer:2011.sph.vs.arepo.shocks}
{Bauer}, A., \& {Springel}, V. 2012, \mnras, 423, 3102

\bibitem[{{Bournaud} {et~al.}(2011){Bournaud}, {Dekel}, {Teyssier}, {Cacciato},
  {Daddi}, {Juneau}, \& {Shankar}}]{bournaud:2011.agn.fueling.by.clumps}
{Bournaud}, F., {Dekel}, A., {Teyssier}, R., {Cacciato}, M., {Daddi}, E.,
  {Juneau}, S., \& {Shankar}, F. 2011, \apjl, 741, L33

\bibitem[{Bournaud {et~al.}(2007)Bournaud, Elmegreen, \&
  Elmegreen}]{bournaud:disk.clumps.to.bulge}
Bournaud, F., Elmegreen, B.~G., \& Elmegreen, D.~M. 2007, The Astrophysical
  Journal, 670, 237

\bibitem[{{Bournaud} {et~al.}(2008)}]{bournaud:chain.gal.model}
{Bournaud}, F., {et~al.} 2008, \aap, 486, 741

\bibitem[{{Ceverino} {et~al.}(2010){Ceverino}, {Dekel}, \&
  {Bournaud}}]{ceverino:2010.clump.disks.cosmosims}
{Ceverino}, D., {Dekel}, A., \& {Bournaud}, F. 2010, \mnras, 404, 2151

\bibitem[{{Ceverino} {et~al.}(2012){Ceverino}, {Dekel}, {Mandelker},
  {Bournaud}, {Burkert}, {Genzel}, \& {Primack}}]{ceverino:2012.clump.rotation}
{Ceverino}, D., {Dekel}, A., {Mandelker}, N., {Bournaud}, F., {Burkert}, A.,
  {Genzel}, R., \& {Primack}, J. 2012, \mnras, 420, 3490

\bibitem[{{Chapman} {et~al.}(2005){Chapman}, {Blain}, {Smail}, \&
  {Ivison}}]{chapman:submm.lfs}
{Chapman}, S.~C., {Blain}, A.~W., {Smail}, I., \& {Ivison}, R.~J. 2005, \apj,
  622, 772

\bibitem[{{Cowie} {et~al.}(1995){Cowie}, {Hu}, \&
  {Songaila}}]{cowie:1995.hst.clump.gal}
{Cowie}, L.~L., {Hu}, E.~M., \& {Songaila}, A. 1995, \aj, 110, 1576

\bibitem[{{Daddi} {et~al.}(2010)}]{daddi:highz.gal.high.fgas}
{Daddi}, E., {et~al.} 2010, \apj, 713, 686

\bibitem[{{Dekel} {et~al.}(2009{\natexlab{a}}){Dekel}, {Sari}, \&
  {Ceverino}}]{dekel:2009.clumpy.disk.evolution.toymodel}
{Dekel}, A., {Sari}, R., \& {Ceverino}, D. 2009{\natexlab{a}}, \apj, 703, 785

\bibitem[{{Dekel} {et~al.}(2009{\natexlab{b}})}]{dekel:cold.streams}
{Dekel}, A., {et~al.} 2009{\natexlab{b}}, \nat, 457, 451

\bibitem[{{Dobbs} {et~al.}(2011){Dobbs}, {Burkert}, \&
  {Pringle}}]{dobbs:2011.why.gmcs.unbound}
{Dobbs}, C.~L., {Burkert}, A., \& {Pringle}, J.~E. 2011, \mnras, 413, 528

\bibitem[{{Elmegreen} {et~al.}(2008){Elmegreen}, {Bournaud}, \&
  {Elmegreen}}]{elmegreen:classical.bulges.from.clumps}
{Elmegreen}, B.~G., {Bournaud}, F., \& {Elmegreen}, D.~M. 2008, \apj, 688, 67

\bibitem[{{Elmegreen} {et~al.}(2009){Elmegreen}, {Elmegreen}, {Fernandez}, \&
  {Lemonias}}]{elmegreen:2009.clump.properties.hudf}
{Elmegreen}, B.~G., {Elmegreen}, D.~M., {Fernandez}, M.~X., \& {Lemonias},
  J.~J. 2009, \apj, 692, 12

\bibitem[{{Elmegreen} {et~al.}(2005{\natexlab{a}}){Elmegreen}, {Elmegreen},
  {Vollbach}, {Foster}, \&
  {Ferguson}}]{elmegreen:2005.exponential.disk.by.clump.form.hiz.small}
{Elmegreen}, B.~G., {Elmegreen}, D.~M., {Vollbach}, D.~R., {Foster}, E.~R., \&
  {Ferguson}, T.~E. 2005{\natexlab{a}}, \apj, 634, 101

\bibitem[{{Elmegreen} \&
  {Elmegreen}(2006)}]{elmegreen:2006.rings.as.linblad.for.clumps}
{Elmegreen}, D.~M., \& {Elmegreen}, B.~G. 2006, \apj, 651, 676

\bibitem[{{Elmegreen} {et~al.}(2004{\natexlab{a}}){Elmegreen}, {Elmegreen}, \&
  {Hirst}}]{elmegreen:2004.chain.gal.faceon}
{Elmegreen}, D.~M., {Elmegreen}, B.~G., \& {Hirst}, A.~C. 2004{\natexlab{a}},
  \apjl, 604, L21

\bibitem[{{Elmegreen} {et~al.}(2005{\natexlab{b}}){Elmegreen}, {Elmegreen},
  {Rubin}, \& {Schaffer}}]{elmegreen:2005.hudf.morphologies.clumps}
{Elmegreen}, D.~M., {Elmegreen}, B.~G., {Rubin}, D.~S., \& {Schaffer}, M.~A.
  2005{\natexlab{b}}, \apj, 631, 85

\bibitem[{{Elmegreen} {et~al.}(2004{\natexlab{b}}){Elmegreen}, {Elmegreen}, \&
  {Sheets}}]{elmegreen:2004.chain.gal}
{Elmegreen}, D.~M., {Elmegreen}, B.~G., \& {Sheets}, C.~M. 2004{\natexlab{b}},
  \apj, 603, 74

\bibitem[{{Evans} {et~al.}(2009)}]{evans:2009.sf.efficiencies.lifetimes}
{Evans}, N.~J., {et~al.} 2009, \apjs, 181, 321

\bibitem[{{Evans}(1999)}]{evans:1999.sf.gmc.review}
{Evans}, II, N.~J. 1999, \araa, 37, 311

\bibitem[{{Forbes} {et~al.}(2012){Forbes}, {Krumholz}, \&
  {Burkert}}]{forbes:2011.thick.disk.torque.evol}
{Forbes}, J., {Krumholz}, M.~R., \& {Burkert}, A. 2012, \apj, 754, 48

\bibitem[{{F{\"o}rster Schreiber}
  {et~al.}(2006)}]{forsterschreiber:z2.disk.turbulence}
{F{\"o}rster Schreiber}, N.~M., {et~al.} 2006, \apj, 645, 1062

\bibitem[{{F{\"o}rster Schreiber}
  {et~al.}(2011{\natexlab{a}})}]{forster-schreiber:2011.hiz.gal.morph}
---. 2011{\natexlab{a}}, \apj, 731, 65

\bibitem[{{F{\"o}rster Schreiber}
  {et~al.}(2011{\natexlab{b}})}]{forster-schreiber:2011.ifu.clump.optical.obs}
---. 2011{\natexlab{b}}, \apj, 739, 45

\bibitem[{{Gammie}(2001)}]{gammie:2001.cooling.in.keplerian.disks}
{Gammie}, C.~F. 2001, \apj, 553, 174

\bibitem[{{Genel} {et~al.}(2012)}]{genel10}
{Genel}, S., {et~al.} 2012, \apj, 745, 11

\bibitem[{{Genzel} {et~al.}(2008)}]{genzel:highz.rapid.secular}
{Genzel}, R., {et~al.} 2008, \apj, 687, 59

\bibitem[{{Genzel} {et~al.}(2011)}]{genzel2011:clumps}
---. 2011, \apj, 733, 101

\bibitem[{{Giavalisco} {et~al.}(1996){Giavalisco}, {Steidel}, \&
  {Macchetto}}]{giavalisco:1996.hst.clump.gal}
{Giavalisco}, M., {Steidel}, C.~C., \& {Macchetto}, F.~D. 1996, \apj, 470, 189

\bibitem[{{Governato} {et~al.}(2004)}]{governato04:resolution.fx}
{Governato}, F., {et~al.} 2004, \apj, 607, 688

\bibitem[{{Governato} {et~al.}(2007)}]{governato:disk.formation}
---. 2007, \mnras, 374, 1479

\bibitem[{{Greve} {et~al.}(2005)}]{greve:smg.co.properties.vs.ell.models}
{Greve}, T.~R., {et~al.} 2005, \mnras, 359, 1165

\bibitem[{{Griffiths} {et~al.}(1994)}]{griffiths:1994.hst.clump.gal}
{Griffiths}, R.~E., {et~al.} 1994, \apjl, 435, L19

\bibitem[{{Hammer} {et~al.}(2009){Hammer}, {Flores}, {Puech}, {Yang},
  {Athanassoula}, {Rodrigues}, \&
  {Delgado}}]{hammer:hubble.sequence.vs.mergers}
{Hammer}, F., {Flores}, H., {Puech}, M., {Yang}, Y.~B., {Athanassoula}, E.,
  {Rodrigues}, M., \& {Delgado}, R. 2009, \aap, 507, 1313

\bibitem[{{Hayward} {et~al.}(2011)}]{hayward:arepo.gadget.mergers}
{Hayward}, C.~C., {et~al.} 2011, \mnras, in preparation

\bibitem[{{Hernquist}(1989)}]{hernquist.89}
{Hernquist}, L. 1989, \nat, 340, 687

\bibitem[{{Hernquist}(1990)}]{hernquist:profile}
---. 1990, \apj, 356, 359

\bibitem[{{Hopkins}(2012)}]{hopkins:excursion.ism}
{Hopkins}, P.~F. 2012, \mnras, 423, 2016

\bibitem[{{Hopkins} {et~al.}(2010){Hopkins}, {Bundy}, {Croton}, {Hernquist},
  {Keres}, {Khochfar}, {Stewart}, {Wetzel}, \&
  {Younger}}]{hopkins:merger.rates}
{Hopkins}, P.~F., {Bundy}, K., {Croton}, D., {Hernquist}, L., {Keres}, D.,
  {Khochfar}, S., {Stewart}, K., {Wetzel}, A., \& {Younger}, J.~D. 2010, \apj,
  715, 202

\bibitem[{{Hopkins} {et~al.}(2009){Hopkins}, {Cox}, {Dutta}, {Hernquist},
  {Kormendy}, \& {Lauer}}]{hopkins:cusps.ell}
{Hopkins}, P.~F., {Cox}, T.~J., {Dutta}, S.~N., {Hernquist}, L., {Kormendy},
  J., \& {Lauer}, T.~R. 2009, \apjs, 181, 135

\bibitem[{{Hopkins} \& {Hernquist}(2010)}]{hopkins:sb.ir.lfs}
{Hopkins}, P.~F., \& {Hernquist}, L. 2010, \mnras, 402, 985

\bibitem[{{Hopkins} {et~al.}(2005){Hopkins}, {Hernquist}, {Martini}, {Cox},
  {Robertson}, {Di Matteo}, \& {Springel}}]{hopkins:lifetimes.letter}
{Hopkins}, P.~F., {Hernquist}, L., {Martini}, P., {Cox}, T.~J., {Robertson},
  B., {Di Matteo}, T., \& {Springel}, V. 2005, \apjl, 625, L71

\bibitem[{{Hopkins} \& {Quataert}(2011)}]{hopkins:inflow.analytics}
{Hopkins}, P.~F., \& {Quataert}, E. 2011, \mnras, 415, 1027

\bibitem[{{Hopkins} {et~al.}(2011){Hopkins}, {Quataert}, \&
  {Murray}}]{hopkins:rad.pressure.sf.fb}
{Hopkins}, P.~F., {Quataert}, E., \& {Murray}, N. 2011, \mnras, 417, 950

\bibitem[{{Hopkins} {et~al.}(2012{\natexlab{a}}){Hopkins}, {Quataert}, \&
  {Murray}}]{hopkins:stellar.fb.winds}
---. 2012{\natexlab{a}}, \mnras, 421, 3522

\bibitem[{{Hopkins} {et~al.}(2012{\natexlab{b}}){Hopkins}, {Quataert}, \&
  {Murray}}]{hopkins:fb.ism.prop}
---. 2012{\natexlab{b}}, \mnras, 421, 3488

\bibitem[{Immeli {et~al.}(2004)Immeli, Samland, Westera, \&
  Gerhard}]{immeli:fragmentation.vs.clump.properties}
Immeli, A., Samland, M., Westera, P., \& Gerhard, O. 2004, The Astrophysical
  Journal, 611, 20

\bibitem[{{Kere{\v s}} {et~al.}(2005){Kere{\v s}}, {Katz}, {Weinberg}, \&
  {Dav{\'e}}}]{keres:hot.halos}
{Kere{\v s}}, D., {Katz}, N., {Weinberg}, D.~H., \& {Dav{\'e}}, R. 2005,
  \mnras, 363, 2

\bibitem[{{Kere{\v s}} {et~al.}(2012){Kere{\v s}}, {Vogelsberger}, {Sijacki},
  {Springel}, \& {Hernquist}}]{keres:2011.arepo.gadget.disk.angmom}
{Kere{\v s}}, D., {Vogelsberger}, M., {Sijacki}, D., {Springel}, V., \&
  {Hernquist}, L. 2012, \mnras, 425, 2027

\bibitem[{{Kriek} {et~al.}(2009){Kriek}, {van Dokkum}, {Franx}, {Illingworth},
  \& {Magee}}]{kriek:z2.hubble.sequence}
{Kriek}, M., {van Dokkum}, P.~G., {Franx}, M., {Illingworth}, G.~D., \&
  {Magee}, D.~K. 2009, \apjl, 705, L71

\bibitem[{{Kroupa}(2002)}]{kroupa:imf}
{Kroupa}, P. 2002, Science, 295, 82

\bibitem[{{Krumholz} \& {Dekel}(2010)}]{krumholz:2010.highz.clump.survival}
{Krumholz}, M.~R., \& {Dekel}, A. 2010, \mnras, 406, 112

\bibitem[{{Krumholz} {et~al.}(2012){Krumholz}, {Dekel}, \&
  {McKee}}]{krumholz:2012.universal.sf.efficiency}
{Krumholz}, M.~R., {Dekel}, A., \& {McKee}, C.~F. 2012, \apj, 745, 69

\bibitem[{{Krumholz} \& {Gnedin}(2011)}]{krumholz:2011.molecular.prescription}
{Krumholz}, M.~R., \& {Gnedin}, N.~Y. 2011, \apj, 729, 36

\bibitem[{{Krumholz} \& {Tan}(2007)}]{krumholz:sf.eff.in.clouds}
{Krumholz}, M.~R., \& {Tan}, J.~C. 2007, \apj, 654, 304

\bibitem[{{Krumholz} \&
  {Thompson}(2012)}]{krumholz:2012.rad.pressure.rt.instab}
{Krumholz}, M.~R., \& {Thompson}, T.~A. 2012, \apj, submitted, arXiv:1203.2926

\bibitem[{{Kuiper} {et~al.}(2012){Kuiper}, {Klahr}, {Beuther}, \&
  {Henning}}]{kuiper:2012.rad.pressure.outflow.vs.rt.method}
{Kuiper}, R., {Klahr}, H., {Beuther}, H., \& {Henning}, T. 2012, \aap, 537,
  A122

\bibitem[{{Leitherer} {et~al.}(1999)}]{starburst99}
{Leitherer}, C., {et~al.} 1999, \apjs, 123, 3

\bibitem[{{Mac Low} \& {Klessen}(2004)}]{mac-low:2004.turb.sf.review}
{Mac Low}, M.-M., \& {Klessen}, R.~S. 2004, Reviews of Modern Physics, 76, 125

\bibitem[{{Mannucci} {et~al.}(2006){Mannucci}, {Della Valle}, \&
  {Panagia}}]{mannucci:2006.snIa.rates}
{Mannucci}, F., {Della Valle}, M., \& {Panagia}, N. 2006, \mnras, 370, 773

\bibitem[{{Murray} {et~al.}(2010){Murray}, {Quataert}, \&
  {Thompson}}]{murray:molcloud.disrupt.by.rad.pressure}
{Murray}, N., {Quataert}, E., \& {Thompson}, T.~A. 2010, \apj, 709, 191

\bibitem[{{Newman} {et~al.}(2012)}]{newman:z2.clump.winds}
{Newman}, S.~F., {et~al.} 2012, \apj, 752, 111

\bibitem[{{Noguchi}(1999)}]{noguchi:1999.clumpy.disk.bulge.formation}
{Noguchi}, M. 1999, \apj, 514, 77

\bibitem[{{Oppenheimer} {et~al.}(2010){Oppenheimer}, {Dav{\'e}}, {Kere{\v s}},
  {Fardal}, {Katz}, {Kollmeier}, \&
  {Weinberg}}]{oppenheimer:recycled.wind.accretion}
{Oppenheimer}, B.~D., {Dav{\'e}}, R., {Kere{\v s}}, D., {Fardal}, M., {Katz},
  N., {Kollmeier}, J.~A., \& {Weinberg}, D.~H. 2010, \mnras, 406, 2325

\bibitem[{{Overzier} {et~al.}(2009)}]{overzier:local.lbgs.w.clumps.vs.mergers}
{Overzier}, R.~A., {et~al.} 2009, \apj, 706, 203

\bibitem[{{Overzier} {et~al.}(2010)}]{overzier:lbgs.clumpy.disks.are.mergers}
---. 2010, \apj, 710, 979

\bibitem[{{Pei}(1992)}]{pei92:reddening.curves}
{Pei}, Y.~C. 1992, \apj, 395, 130

\bibitem[{{Petty} {et~al.}(2009){Petty}, {de Mello}, {Gallagher}, {Gardner},
  {Lotz}, {Matt Mountain}, \&
  {Smith}}]{petty:local.clumpy.systems.same.as.highz}
{Petty}, S.~M., {de Mello}, D.~F., {Gallagher}, J.~S., {Gardner}, J.~P.,
  {Lotz}, J.~M., {Matt Mountain}, C., \& {Smith}, L.~J. 2009, \aj, 138, 362

\bibitem[{{Planesas} {et~al.}(1991){Planesas}, {Scoville}, \&
  {Myers}}]{planesas:1991.1068.superclumps}
{Planesas}, P., {Scoville}, N., \& {Myers}, S.~T. 1991, \apj, 369, 364

\bibitem[{{Rand} \& {Kulkarni}(1990)}]{rand:1990.m51.superclumps}
{Rand}, R.~J., \& {Kulkarni}, S.~R. 1990, \apjl, 349, L43

\bibitem[{{Robertson} {et~al.}(2004){Robertson}, {Yoshida}, {Springel}, \&
  {Hernquist}}]{robertson:cosmological.disk.formation}
{Robertson}, B., {Yoshida}, N., {Springel}, V., \& {Hernquist}, L. 2004, \apj,
  606, 32

\bibitem[{{Robertson} \&
  {Bullock}(2008)}]{robertsonbullock:highz.disk.vs.model}
{Robertson}, B.~E., \& {Bullock}, J.~S. 2008, \apjl, 685, L27

\bibitem[{{Saitoh} \& {Makino}(2012)}]{saitoh:2012.dens.indep.sph}
{Saitoh}, T.~R., \& {Makino}, J. 2012, \apj, in press, arXiv:1202.4277

\bibitem[{Shapiro {et~al.}(2009)}]{shapiro:broad.halpha.emission.highz.gal}
Shapiro, K., {et~al.} 2009, \apj, 701, 955, 11 pages, 6 figures; submitted to
  ApJ

\bibitem[{{Shapiro} {et~al.}(2008)}]{shapiro:highz.kinematics}
{Shapiro}, K.~L., {et~al.} 2008, \apj, 682, 231

\bibitem[{{Shlosman} \&
  {Noguchi}(1993)}]{shlosman:1993.clumpy.disk.instab.sims}
{Shlosman}, I., \& {Noguchi}, M. 1993, \apj, 414, 474

\bibitem[{{Sijacki} {et~al.}(2012){Sijacki}, {Vogelsberger}, {Keres},
  {Springel}, \& {Hernquist}}]{sijacki:2011.gadget.arepo.hydro.tests}
{Sijacki}, D., {Vogelsberger}, M., {Keres}, D., {Springel}, V., \& {Hernquist},
  L. 2012, \mnras, 424, 2999

\bibitem[{{Somerville} {et~al.}(2001){Somerville}, {Primack}, \&
  {Faber}}]{somerville:sam}
{Somerville}, R.~S., {Primack}, J.~R., \& {Faber}, S.~M. 2001, \mnras, 320, 504

\bibitem[{{Sommer-Larsen} {et~al.}(1999){Sommer-Larsen}, {Gelato}, \&
  {Vedel}}]{sommerlarsen99:disk.sne.fb}
{Sommer-Larsen}, J., {Gelato}, S., \& {Vedel}, H. 1999, \apj, 519, 501

\bibitem[{{Springel}(2005)}]{springel:gadget}
{Springel}, V. 2005, \mnras, 364, 1105

\bibitem[{Springel(2010)}]{springel:arepo}
Springel, V. 2010, \mnras, 401, 791

\bibitem[{{Springel} \& {Hernquist}(2003)}]{springel:multiphase}
{Springel}, V., \& {Hernquist}, L. 2003, \mnras, 339, 289

\bibitem[{{Springel} {et~al.}(2001){Springel}, {White}, {Tormen}, \&
  {Kauffmann}}]{springel:cluster.subhalos}
{Springel}, V., {White}, S.~D.~M., {Tormen}, G., \& {Kauffmann}, G. 2001,
  \mnras, 328, 726

\bibitem[{{Stewart} {et~al.}(2009){Stewart}, {Bullock}, {Barton}, \&
  {Wechsler}}]{stewart:merger.rates}
{Stewart}, K.~R., {Bullock}, J.~S., {Barton}, E.~J., \& {Wechsler}, R.~H. 2009,
  \apj, 702, 1005

\bibitem[{{Swinbank} {et~al.}(2011){Swinbank}, {Papadopoulos}, {Cox}, {Krips},
  {Ivison}, {Smail}, {Thomson}, {Neri}, {Richard}, \&
  {Ebeling}}]{swinbank:clumps}
{Swinbank}, A.~M., {Papadopoulos}, P.~P., {Cox}, P., {Krips}, M., {Ivison},
  R.~J., {Smail}, I., {Thomson}, A.~P., {Neri}, R., {Richard}, J., \&
  {Ebeling}, H. 2011, \apj, 742, 11

\bibitem[{{Tacconi} {et~al.}(2006)}]{tacconi:smg.maximal.sb.sizes}
{Tacconi}, L.~J., {et~al.} 2006, \apj, 640, 228

\bibitem[{{Tacconi} {et~al.}(2008)}]{tacconi:smg.mgr.lifetime.to.quiescent}
---. 2008, \apj, 680, 246

\bibitem[{{Tacconi} {et~al.}(2010)}]{tacconi:high.molecular.gf.highz}
---. 2010, \nat, 463, 781

\bibitem[{{Tasker}(2011)}]{tasker:2011.photoion.heating.gmc.evol}
{Tasker}, E.~J. 2011, \apj, 730, 11

\bibitem[{{Toomre}(1964)}]{toomre:Q}
{Toomre}, A. 1964, \apj, 139, 1217

\bibitem[{{Torrey} {et~al.}(2011){Torrey}, {Vogelsberger}, {Sijacki},
  {Springel}, \& {Hernquist}}]{torrey:2011.arepo.disks}
{Torrey}, P., {Vogelsberger}, M., {Sijacki}, D., {Springel}, V., \&
  {Hernquist}, L. 2011, \apj, in press, arXiv:1110.5635

\bibitem[{{Vogelsberger} {et~al.}(2011){Vogelsberger}, {Sijacki}, {Keres},
  {Springel}, \& {Hernquist}}]{vogelsberger:2011.arepo.vs.gadget.cosmo}
{Vogelsberger}, M., {Sijacki}, D., {Keres}, D., {Springel}, V., \& {Hernquist},
  L. 2011, \mnras, in press arXiv:1109.1281

\bibitem[{{Williams} \& {McKee}(1997)}]{williams:1997.gmc.prop}
{Williams}, J.~P., \& {McKee}, C.~F. 1997, \apj, 476, 166

\bibitem[{{Wilson} {et~al.}(2006){Wilson}, {Harris}, {Longden}, \&
  {Scoville}}]{wilson:2006.arp220.superclumps}
{Wilson}, C.~D., {Harris}, W.~E., {Longden}, R., \& {Scoville}, N.~Z. 2006,
  \apj, 641, 763

\bibitem[{{Wilson} {et~al.}(2003){Wilson}, {Scoville}, {Madden}, \&
  {Charmandaris}}]{wilson:2003.supergiant.clumps}
{Wilson}, C.~D., {Scoville}, N., {Madden}, S.~C., \& {Charmandaris}, V. 2003,
  \apj, 599, 1049

\bibitem[{{Wuyts} {et~al.}(2012)}]{wuyts:2012.no.clumps.in.candels.stellarmass}
{Wuyts}, S., {et~al.} 2012, \apj, 753, 114

\bibitem[{{Zuckerman} \& {Evans}(1974)}]{zuckerman:1974.gmc.constraints}
{Zuckerman}, B., \& {Evans}, II, N.~J. 1974, \apjl, 192, L149

\end{thebibliography}

\begin{appendix}

\section{Numerical Effects of the Radiation Pressure Implementation}

The numerical details and tests of the feedback models are discussed in great detail in \paperone, \papertwo, and \paperthree. However, since we argue that radiation pressure is the most important element of feedback in these dense systems, we here discuss how our main conclusions change if we vary numerical aspects of the radiation pressure model implementation.

\begin{figure}
    \centering
    \plotonesize{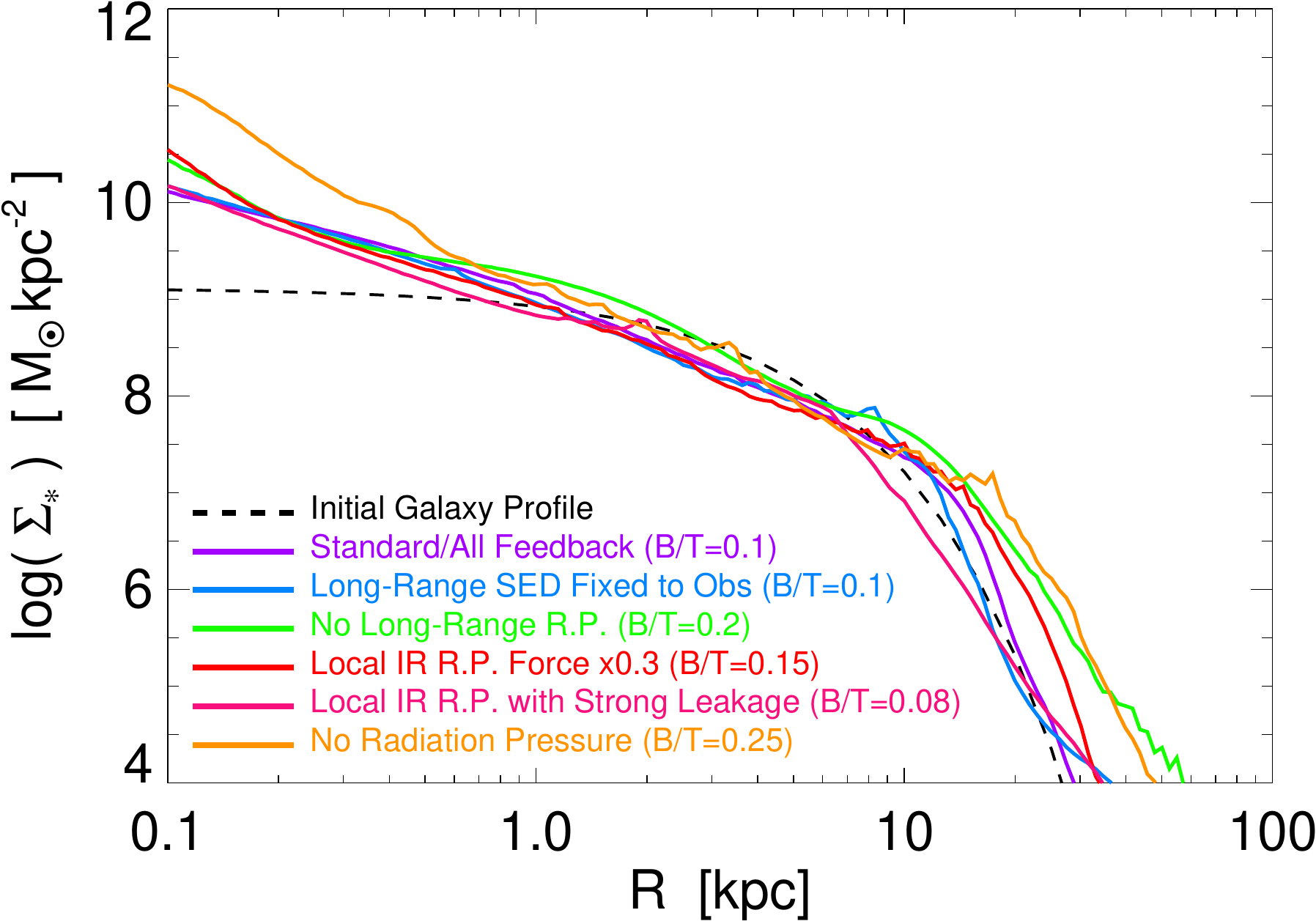}{1.0}
    \caption{Remnant stellar mass profiles as Fig.~\ref{fig:massprofile}; here we specifically compare models where we freely vary sub-elements of the radiation pressure implementation. 
    The ``standard/all feedback'' model and ``no radiation pressure model'' are identical to Fig.~\ref{fig:massprofile}. 
    The ``long-range SED fixed to obs.'' model fixes the spectral shape of the propagated long-range radiation pressure (RP) terms (mechanism [5] in \S~\ref{sec:sims}) to the mean observed in the galaxies modeled. This has little effect, since the self-consistently calculated SEDs agree well with the observations
    The ``no long-range RP'' model removes the long-range terms entirely. This significantly suppresses the mixing and disruption of the largest super-clump complexes, resulting in more bulge growth and internal evolution. 
    The ``local IR RP force x$0.3$'' model suppresses the local IR RP term (mechanism [1] in \S~\ref{sec:sims}) by an arbitrary factor $=0.3$. This leads to less feedback-regulation in the dense gas and greater bulge growth, but the effect is weak when the long-range RP is still present.
    The ``local IR RP with strong leakage'' model calculates the local IR RP term assuming a sub-grid fractal ISM with a broad density PDF giving a much larger UV/optical escape fraction and weaker multiple-scattering ``boost'' for a given average dust optical depth in the region. Although this makes the local RP term weaker, the enhanced escape of UV/optical photons from the very dense regions around young stars makes the long-range RP forces significantly stronger and increases the coherent acceleration of gas between the dense sub-cloud cores in the massive cloud complexes (as well as the launching of RP-driven winds); as a result this slightly suppresses the internal evolution relative to the standard model. 
    \label{fig:profile.appendix}}
\end{figure}

In Fig.~\ref{fig:profile.appendix}, we plot the final mass profiles in the simulations as in Fig.~\ref{fig:massprofile}, but vary the radiation pressure physics included. The ``standard'' (all feedback mechanisms included) case and ``no radiation pressure'' case are the same as Fig.~\ref{fig:massprofile}. 

Recall, the radiation pressure force is coupled in two separate algorithms, the ``local IR'' and ``long-range'' terms described in \S~\ref{sec:sims} (mechanisms [1] and [5], respectively). The former includes the effect of primary optical/UV photons absorbed in the immediate vicinity of young stars (within of order a smoothing length), as well as multiple-scattering of the re-emitted IR photons when the IR optical depth $\tau$ is significant (which dominate this term when $\tau\gtrsim1$). The latter includes the effect of the photons escaping this region (UV/optical, and IR) as they propagate to infinity; although the IR term is included, since most of the volume outside the dense regions near stars is optically thin to it, most of the ``work'' in this term is done by the UV/optical photons being absorbed by less-dense gas further from the dense regions of origin. It is instructive to decompose the effects of these terms. 

In the Appendix of \papertwo, we discuss at length the accuracy of our long-range radiation pressure approximation. There, we show that although we are not solving the full radiation hydrodynamic equations, the simplifications used here are a surprisingly good approximation to full radiative transfer solutions. Nevertheless, it is important to check how big an effect this has. In Fig.~\ref{fig:profile.appendix}, we therefore re-run our simulation, but instead of using the on-the-fly simulation information on stellar populations and gas distributions to calculate the spectral shape escaping and propagating in the long-range radiation pressure term, we simply fix the SED shape to one chosen to match observations. The details of this procedure are given in the Appendix of \papertwo; as noted there, choosing instead a mean SED matched to a full radiative transfer calculation on the simulations gives an almost identical result (since it closely matches what is observed). This has almost no effect on our conclusions relative to our ``standard'' model. 

If we remove the long-range terms (mechanism [5] in \S~\ref{sec:sims}) entirely, but still include the local/IR term (mechanism [1]), we see a much more significant difference. Both the bulge mass and internal evolution of the outer disk are significantly enhanced above the ``standard'' model. Most of the new bulge mass is at $\sim1\,$kpc. The very central regions at $\lesssim100\,$pc are regulated by the IR term and so are still maintained at low densities, but the overall ability to disrupt the $\sim$kpc-scale clump complexes is reduced. The net effect is comparable to -- although still clearly less dramatic than -- the effects of removing radiation pressure entirely. 

It therefore seems that the long-range radiation acceleration is at least comparable, if not more important, to the local IR radiation pressure in suppressing the clump star formation efficiencies and subsequent bulge growth. This is true for several reasons: but the critical point is that the kpc-scale complexes have clear substructure. As shown in \papertwo, at any given instant, most of the mass in complexes is not in the extremely dense sub-regions forming stars, and this ``intermediate-density'' material is essentially {\em never} tightly bound or virialized inside a cloud when feedback is present. The IR radiation pressure, obviously, acts significantly only where the IR opacities are substantial. The volume-average surface densities of the complexes are only marginally optically thick to the IR ($\tau_{\rm IR}\sim1$ for $\Sigma_{\rm gas}\sim 10^{9}\,\msun\,{\rm kpc^{-2}}$); so the IR coupling is negligible in the lower-density volume-filling medium inside each complex. The IR coupling is therefore dominated (as shown in \paperone) by very dense sub-clumps at $\sim1\,$pc scales (densities $\Sigma_{\rm gas}\gg 10^{4}\,\msun\,{\rm kpc^{-2}}$ and three-dimensional densities $n\gtrsim 10^{5}\,{\rm cm^{-3}}$), over a very short time when they have just formed young stars. But these dense sub-regions are distributed throughout the complex, so the IR term does not act {\em coherently} across the entire cloud complex to disrupt it. Rather, it disrupts the dense star-forming sub-regions shortly after they form stars, recycling the dense gas into the parent complex, at turbulent velocities of order the local escape velocity from the sub-clump ($\sim10-50\,{\rm km\,s^{-1}}$). This makes it optically thin to the IR photons, but (being marginally bound) much more vulnerable to continuous, {\em coherent} acceleration from the collective UV/optical flux of the all the stars in the complex. 

The short-range radiation pressure mechanism -- in particular the multiple-scattering of IR photons producing a local momentum flux of $\sim \tau\, L/c$ -- can, in principle, be more affected by the details of accurate radiation transport. For example, \citet{krumholz:2012.rad.pressure.rt.instab} argue that, in a time-averaged sense, the ``boost factor'' $\tau_{e}$ in $\dot{p}\sim \tau_{e}\,L/c$ does not scale linearly with the full average optical depth $\tau$ when $\tau\gg1$ and does not typically exceed a factor of several. We should stress, however, that our conclusions are actually not that different from those of \citet{krumholz:2012.rad.pressure.rt.instab}. Those authors argue specifically that the reason $\tau_{e}$ is ``capped'' is because, whenever the radiation pressure force temporarily exceeds gravity, gas is locally blown away and dense regions are ``shredded'' by radiation Rayleigh-Taylor instabilities, until the coupling is lowered such that radiation balances gravity. Even though we do not follow the full radiation transport, this is the generic behavior in our simulations as well (IR radiation pressure drives to lower effective $\tau_{e}$ to offset gravity). As a result, as discussed in \paperone\ and \papertwo, the IR term primarily drives local turbulence, disrupting or shredding dense sub-regions and driving turbulence to maintain a Toomre $Q\approx1$, while the long-range radiation pressure (which operates via single scattering and is not affected by this argument) drives winds in loosely-bound material, primarily coupling via its UV/optical opacity. Consider, for example, in \paperone, we show that the mean true IR $\tau$ in the dense protostellar core-like regions that have just formed new stars -- which dominate the momentum input -- can reach as high as $\sim10-50$. However, we also show that the coupling of $\tau\,L/c$ in these regions immediately (often in $<1\,$Myr) disperses the gas to lower $\tau$ around the source and allows resolved leakage, so that the {\em luminosity-averaged} coupling factor, i.e.\ the true $\tau_{e}$ we should compare to \citet{krumholz:2012.rad.pressure.rt.instab}, is actually just $\tau_{e}\approx4-5$ in our simulations. Both this and the resulting local velocity dispersions and Eddington ratios are in reasonable agreement with these and other radiative transfer calculations (\citealt{kuiper:2012.rad.pressure.outflow.vs.rt.method}, Jonsson et al.\ in preparation).

Still, one could imagine sub-resolution scale effects that might reduce this local IR coupling. We discuss this extensively in \paperone\ (Appendix B), and in particular consider the case of a fractal ISM continuing to infinitely small scales below the resolved region used to calculate $\tau$; as opposed to a homogenous medium, this will present ``holes'' or ``channels'' along which photons can leak out. For a lognormal or power-law density distribution, we show that in this particular case it is straightforward to analytically calculate the ``effective'' coupling and leakage effects. In all cases where $\tau$ is large, this dramatically enhances the escape fraction of UV and optical photons. However, if the scatter in density contributed strictly below the resolution scale of $\sim1-10\,$pc is weak or modest, $\lesssim 0.5\,$dex, then this has no effect on the dimensional scaling of the momentum imparted by multiply-scattered IR photons (rather it just amounts to a weak normalization correction). For larger scatter it introduces a scaling $\tau_{e}\sim \tau^{\beta}$ where $\beta<1$ is a weakly decreasing power of the sub-resolution density dispersion. In \paperone\ we show how this can be implemented in the simulations and used to allow both for the corrected IR coupling and UV/optical escape fractions. We therefore consider two examples using this ``leakage'' estimator. In the first, we follow \paperone\ (model ``PL: $\sigma$ calculated'' therein) and use the local (resolved) density dispersion added in quadrature with a large assumed floor of $\sim1\,$dex and the solution for a fractal medium -- this results in a highly sub-linear $\tau_{e}(\tau)$ (approximately $\propto \tau^{1/3}$), and actually produces significantly weaker $\tau_{e}$ values than those calculated in full radiative transfer calculations by either \citet{krumholz:2012.rad.pressure.rt.instab} or \citet{kuiper:2012.rad.pressure.outflow.vs.rt.method}. But we see little difference between this and our standard model -- if anything, the bulge is slightly smaller! This occurs because, as discussed above, the single-scattering is critical to actually dispersing the cloud complexes, while the IR largely drives local turbulence and is always driven to approximate equilibrium with gravity. The latter equilibration still occurs, even if the coupling is made weaker (gas collapses to slightly higher densities first, giving higher $\tau$ to obtain the same $\tau_{e}$); this is shown explicitly (along with the lack of strong effects on any of our predicted galaxy properties) in \paperone\ (Figs.~5,\,9,\,10,\,B2). And allowing for leakage of course makes the long-range radiation pressure forces {\em stronger}, by removing fewer of the UV/optical photons in the dense regions near stars where they act less efficiently. 

To clearly distinguish between the effects of decreasing the local IR radiation force and increasing the long-range force, we can artificially multiply the local IR radiation pressure by an arbitrary factor, without increasing the UV/optical leakage. Fig.~\ref{fig:profile.appendix} shows such an experiment, with a reduction of the local IR force by a factor of $3$, which leads to a very modest galaxy-averaged $\tau_{e}\approx2$. As Fig.~\ref{fig:profile.appendix} shows, here we do find more bulge growth and longer-lived clouds than in the previous case, as expected, although the effect is small.

Other effects discussed in \paperone, however, could mean that our implementation actually under-estimates the strength of the radiation pressure terms. For example, if the gas density profiles rise sufficiently steeply around young stars, or if we are under-resolving the maximum densities (hence opacities) to which regions surrounding very young stars collapse, then our simulation-calculated momentum flux could be a substantial under-estimate. \citet{kuiper:2012.rad.pressure.outflow.vs.rt.method} also argue that calculations such as \citet{krumholz:2012.rad.pressure.rt.instab}, which do not follow the exact radiative transfer but adopt a pure flux limited diffusion approximation and see efficient dispersal as described above, can underestimate the true coupling efficiency by an order-of-magnitude. Unsurprisingly, we see further reduction of bulge growth and internal evolution in experiments where we enhance the radiation pressure strength to match these suggestions.

Finally, as shown in \paperone\ and \papertwo, we have run resolution tests with these models up to $\sim10^{9}$ particles and find very good convergence at the standard resolution adopted here; in fact for the bulk properties here we see reasonable convergence even at order-of-magnitude lower particle numbers. 

\end{appendix}

\end{document}